\documentclass[aps,prb,twocolumn,amsfonts,amsmath,amssymb,superscriptaddress]{revtex4}
\usepackage{graphicx}
\usepackage{dcolumn}
\usepackage{bm}% bold math

\usepackage{colordvi}
\usepackage{color}

\usepackage[normalem]{ulem}

\begin{document}
\title{Hole spin dynamics and hole $g$ factor anisotropy in coupled quantum well systems}
\author{C. Gradl}
\affiliation{Institut f\"ur Experimentelle und Angewandte Physik,
Universit\"at Regensburg, D-93040 Regensburg, Germany}
\author{M. Kempf}
\affiliation{Institut f\"ur Experimentelle und Angewandte Physik,
Universit\"at Regensburg, D-93040 Regensburg, Germany}
\author{D.\ Schuh}
\affiliation{Institut f\"ur Experimentelle und Angewandte Physik,
Universit\"at Regensburg, D-93040 Regensburg, Germany}
\author{D.\ Bougeard}
\affiliation{Institut f\"ur Experimentelle und Angewandte Physik,
Universit\"at Regensburg, D-93040 Regensburg, Germany}
\author{R.\ Winkler}
\affiliation{Department of Physics, Northern Illinois University, DeKalb IL 60115, USA}
\author{C.\ Sch\"uller}
\affiliation{Institut f\"ur Experimentelle und Angewandte Physik,
Universit\"at Regensburg, D-93040 Regensburg, Germany}
\author{T.\ Korn}
\email{tobias.korn@physik.uni-regensburg.de}
\affiliation{Institut
f\"ur Experimentelle und Angewandte Physik, Universit\"at
Regensburg, D-93040 Regensburg, Germany}
\date{\today}

\begin{abstract}
Due to its p-like character, the valence band in GaAs-based heterostructures offers rich and complex spin-dependent phenomena. One manifestation is the large anisotropy of Zeeman spin splitting. Using undoped, coupled quantum wells (QWs), we examine this anisotropy by comparing the hole spin dynamics for high- and low-symmetry crystallographic orientations of the QWs. We directly measure the hole $g$ factor via time-resolved Kerr rotation, and for the low-symmetry crystallographic orientations (110) and (113a), we observe a large in-plane anisotropy of the hole $g$ factor,  in good agreement with our theoretical calculations. Using resonant spin amplification, we also observe an anisotropy of the hole spin \emph{dephasing} in the (110)-grown structure, indicating that crystal symmetry may be used to control hole spin dynamics.
\end{abstract}
\maketitle
\section{Introduction}
In recent years, spin dynamics in low-dimensional semiconductor structures, such as quantum wells (QWs) and quantum dots, have attracted significant scientific interest.
A large number of studies have been conducted on two-dimensional electron systems (2DESs) confined in QWs, exploiting the symmetry of the spin-orbit fields to control electron spin dynamics. Here, the choice of growth-axis symmetry allows for suppression of spin dephasing for particular spin orientations in (110)-grown QWs~\cite{DP110,Ohno99_1,Oestreich04_1,Griesbeck12},  cancellation of Rashba and Dresselhaus spin-orbit fields in (111)-grown structures~\cite{Cartoixa05, Marie_111}, or formation of a persistent spin helix state in (001)-grown QWs~\cite{bernevig:236601,Koralek09,Salis_Helix}.
Similar studies using these crystallographic degrees of freedom for two-dimensional hole systems (2DHSs) are lacking, even though large anisotropies of the $g$ factor~\cite{winkler00,Kubisa11} and suppression of hole spin dephasing~\cite{Wu_Hole_12} have been predicted. This is, in part, due to the difficulties associated with p-modulation doping, which requires different growth strategies depending on the crystallographic orientation of a 2DHS~\cite{Heremans_2DHS,gerl05,gerl06}. Conventional undoped QWs are not suitable for low-temperature spin dynamics studies, as rapid photocarrier recombination limits the observation window for optically oriented carriers to less than 100~ps. The complex structure of the valence band also complicates studies of hole spin coherence:
In GaAs-based structures, long-lived hole spin coherence can only be expected if the degeneracy between light-hole (LH) and heavy-hole (HH) bands that is present in the bulk~\cite{hilton02} is lifted by confinement. Yet even in confined systems, there is a pronounced mixing of LH and HH bands for wave vectors $k>0$~\cite{winkler03}. Thus, long-lived hole spin coherence is only observed for low-density 2DHSs at low temperatures~\cite{syperek07,kugler:035325,Korn10}. In these conditions, hole spin dephasing times may rival or even exceed those of conduction-band electrons due to the reduced hyperfine interaction of the p-like holes with surrounding nuclei~\cite{fischer:155329,eble:146601}.
\begin{figure}[h]%
\centering
\includegraphics[width= \linewidth]{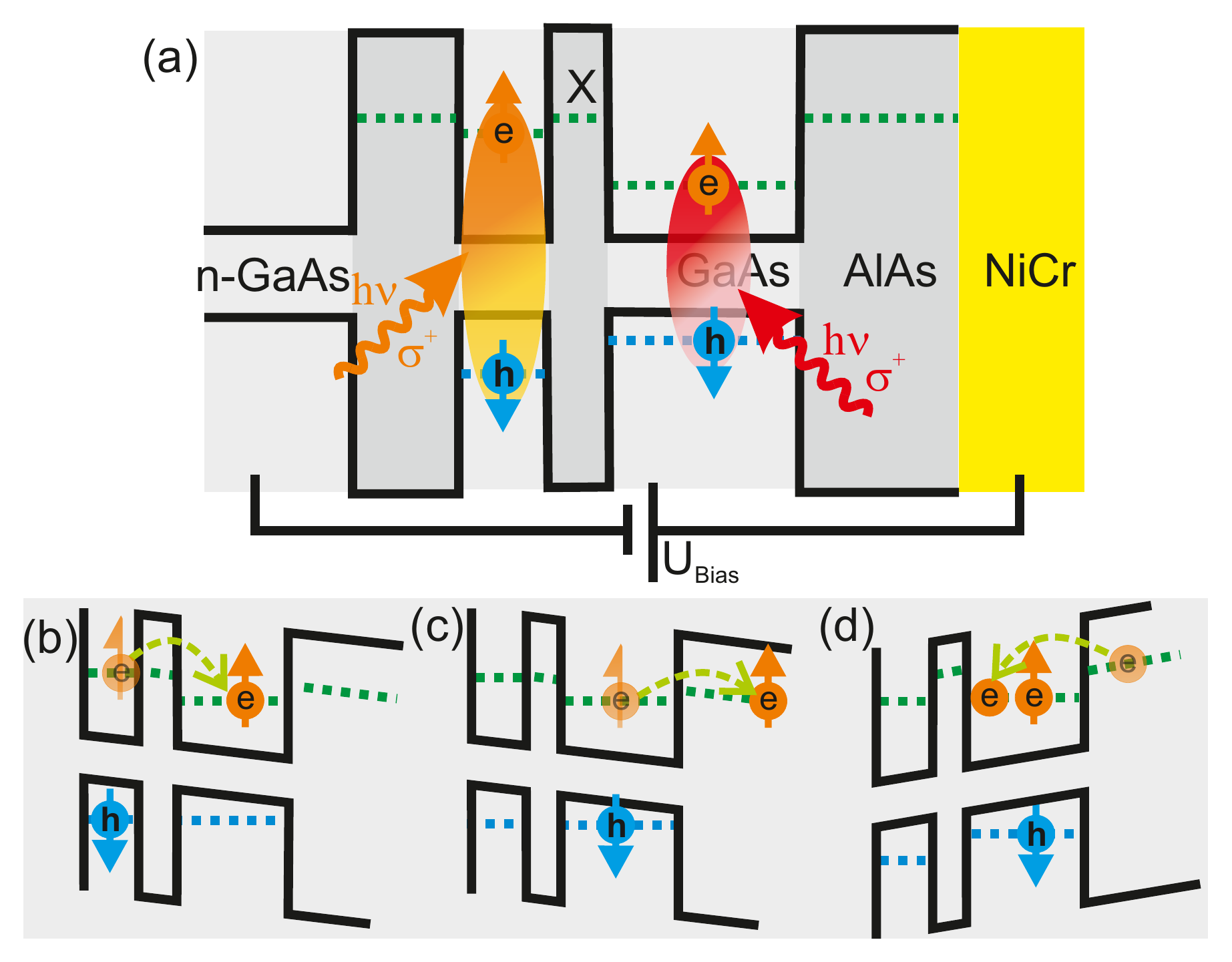}
\caption{(a) Sample structure. Spin-polarized electron-hole pairs can be created by resonant optical excitation in  the narrow or the wide QW.The  states in the X valley of the AlAs barrier layers are energetically close to the electron states in the narrow QW. (b) After resonant excitation in the narrow QW, electrons can rapidly tunnel into the AlAs barrier if the QW is tilted by the gate bias, so that only holes remain in the QW. (c) and (d) Depending on the bias, after  resonant excitation in the wide QW, either electrons tunnel out of the QW, or additional electrons tunnel into the QW from the top contact.}
\label{Fig1}
\end{figure}

To address these challenges, we utilize a special sample design, which is sketched in Fig.~\ref{Fig1}(a). It consists of an \emph{undoped} double QW structure with an AlAs barrier separating a wide and a narrow QW. The  states in the X valley of the AlAs barriers are energetically close to the electron states in the narrow QW, so that fast electron tunneling is possible~\cite{Dawson_Godfrey}. Application of a bias voltage between a back contact and a semi-transparent top gate allows tilting of the QW. By resonant optical excitation,  spin-polarized electron-hole pairs can selectively be created  in the narrow or the wide QW. Figure~\ref{Fig1}(b) depicts resonant excitation of the narrow QW. Optically oriented electrons rapidly tunnel from the narrow QW into the wide QW, while the holes remain in the narrow QW. This spatial separation of the optically oriented electron-hole pairs leads to photocarrier lifetimes of the order of several $\mu$s~\cite{Dawson_Godfrey}, well above the spin dephasing times. By resonant excitation of the wide QW, two different regimes are accessible: depending on the gate bias, additional electrons may tunnel into the QW, or some of the optically oriented electrons may tunnel out, as indicated in Fig.~\ref{Fig1}(c) and (d), creating an imbalance between electron and hole concentrations. After direct photocarrier recombination in the QW, which occurs on a 100~ps timescale,  excess spin-polarized carriers remain in the QW, and their spin precession in an in-plane magnetic field can be observed on a nanosecond timescale.

In the following, we present time-resolved studies of  the spin dynamics for electrons and holes confined in double QW structures grown along different crystallographic orientations.
Three samples, grown on (001) [sample A], (110) [B], and (113a) [C] substrates, are investigated. We demonstrate gate control of the dominant charge carrier type in the QW, and are able to observe long-lived hole spin precession in all samples. The (001)-grown sample serves as a reference to compare hole spin dynamics in our \emph{undoped} double QW structure to previous results obtained on \emph{p-modulation-doped}, (001)-grown samples~\cite{kugler:035325,Korn10,Kugler_nonresonant,Kamil13}.  For the (113a)- and (110)-grown samples, we find a large magnitude, and a pronounced in-plane anisotropy of the hole $g$ factor, in good agreement with our theoretical calculations, and a weaker anisotropy of the electron $g$ factor. In the (110)-grown sample, we additionally observe an  anisotropy of the hole spin dephasing.
\section{Sample structure and experimental methods}
\subsection{Sample design}
All samples were grown by molecular beam epitaxy (MBE) on undoped wafers with different crystallographic orientations [sample A: (001)  B: (110),  C: (113a)]. The active region consists of two GaAs  QWs embedded in AlAs barriers, following a design introduced in Ref.~\onlinecite{Dawson_Godfrey}. A highly n-doped bulk GaAs layer is grown below the active region to serve as a back contact. This layer is contacted from the top by alloying indium contacts. The QWs have nominal widths of 5~nm (narrow QW) and 12~nm (wide QW), respectively, and are separated by an 8~nm wide AlAs barrier. A semitransparent top gate was prepared on the samples. For this, a 10~nm thick SiO$_2$ layer, followed by semitransparent, 6~nm thick NiCr layer, were thermally evaporated on top of the sample.
\subsection{Optical spectroscopy}
A pulsed Ti-sapphire laser system, generating pulses with 2~ps length, and corresponding  spectral width of 1~meV, was used for the time-resolved
measurements. The repetition rate of the laser system is 80~MHz, corresponding to a time delay of 12.5~ns between subsequent
pulses. The laser pulses are split into a circularly-polarized pump beam and a linearly-polarized probe beam by a beam splitter. A
mechanical delay line is used to create a variable time delay between pump and probe. Both beams are focused to a diameter of about
80~$\mu$m on the sample using an achromatic lens. Excitation densities of about 5~Wcm$^{-2}$ are used for the pump beam, corresponding to an optically induced carrier density of about 1.5~$\times 10^{10}$~cm$^{-2}$.

Low-temperature measurements were performed in
an optical cryostat with $^3$He insert where the samples are cooled  to about 1.3~K.  Magnetic fields of up to 11.5~T can be applied, either in the QW plane or perpendicular to it.

In the TRKR and RSA experiments, the circularly-polarized pump beam is generating electron-hole pairs in the QW, with spins aligned parallel or antiparallel to the beam direction, i.e., the QW normal,
depending on the helicity of the light.  In the TRKR measurements, the spin polarization created perpendicular to the sample plane by the pump beam is probed by the time-delayed probe beam via the Kerr effect: the axis of linear polarization of the probe beam is rotated by a small angle, which is proportional to the out-of-plane component of the spin polarization.   This small angle is detected using an optical bridge. A lock-in scheme is used to increase sensitivity. The RSA technique is based on the interference of spin polarizations created in a sample by subsequent pump pulses.  For certain magnetic fields applied in the sample plane, the optically oriented spin polarization precesses by an integer multiple of $2\pi$ in the time window between subsequent pump pulses, so that constructive interference occurs. This leads to pronounced maxima in the  Kerr rotation angle, measured for a fixed time delay as a function of the magnetic field.
\section{Numerical Calculations}
The numerical calculations of Zeeman spin splitting have been based
on the $8\times 8$ Kane Hamiltonian, including the lowest conduction
band $\Gamma_6^c$ as well as the highest valence band $\Gamma_8^v$
and the spin split-off valence band $\Gamma_7^v$ \cite{winkler03}.
The in-plane magnetic field $\bm{B} = (B_x,B_y,0)$ was taken into
account via the vector potential $\bm{A}$ using the asymmetric gauge
$\bm{A}(z) = (zB_y, -z B_x, 0)$.  Diagonalizing the Kane Hamiltonian as
a function of the kinetic momentum $\hbar\bm{k} + e\bm{A}$ then
yields the Zeeman-split energy dispersions $E_{\nu\sigma} (\bm{k})$
of the spin subbands ${\nu}{\uparrow}$ and ${\nu}{\downarrow}$ in
the presence of the magnetic field $\bm{B}$.  This model contains
the $g$ factor only implicitly.  We extract $g$ for the lowest HH
subband $\nu=0$ from the Zeeman splitting
$\Delta E = E_{0\uparrow}(\bm{k}=0) - E_{0\downarrow}(\bm{k}=0)$
calculated at $B = 1$~T using $g = \Delta E/(\mu_\mathrm{B} B)$,
where $\mu_\mathrm{B}$ is the Bohr magneton.  We note that the
simplified expressions for the anisotropic $g$ previously presented
in Ref.~\onlinecite{winkler00} were based on the smaller, less
accurate Luttinger Hamiltonian containing only the highest valence
band $\Gamma_8^v$.  The Luttinger model is thus best suited to
describe hole systems in wide QWs where the confinement energies are
small.  The more accurate Kane model used here contains nonparabolic
corrections to all orders in the wave vector $\bm{k}$ so that it is
appropriate also for the more narrow samples studied here.  All
band-structure parameters entering the Kane model are well-known
from many independent experiments \cite{winkler03}.  Therefore, the
calculations presented here can be regarded as
parameter-free. Numerical algorithms follow Ref.~\onlinecite{winkler03}.
\section{Results and discussion}
\subsection{Electron and hole spin dynamics in (001)-grown coupled QW system}
\begin{figure}[h]%
\centering
\includegraphics[width= \linewidth]{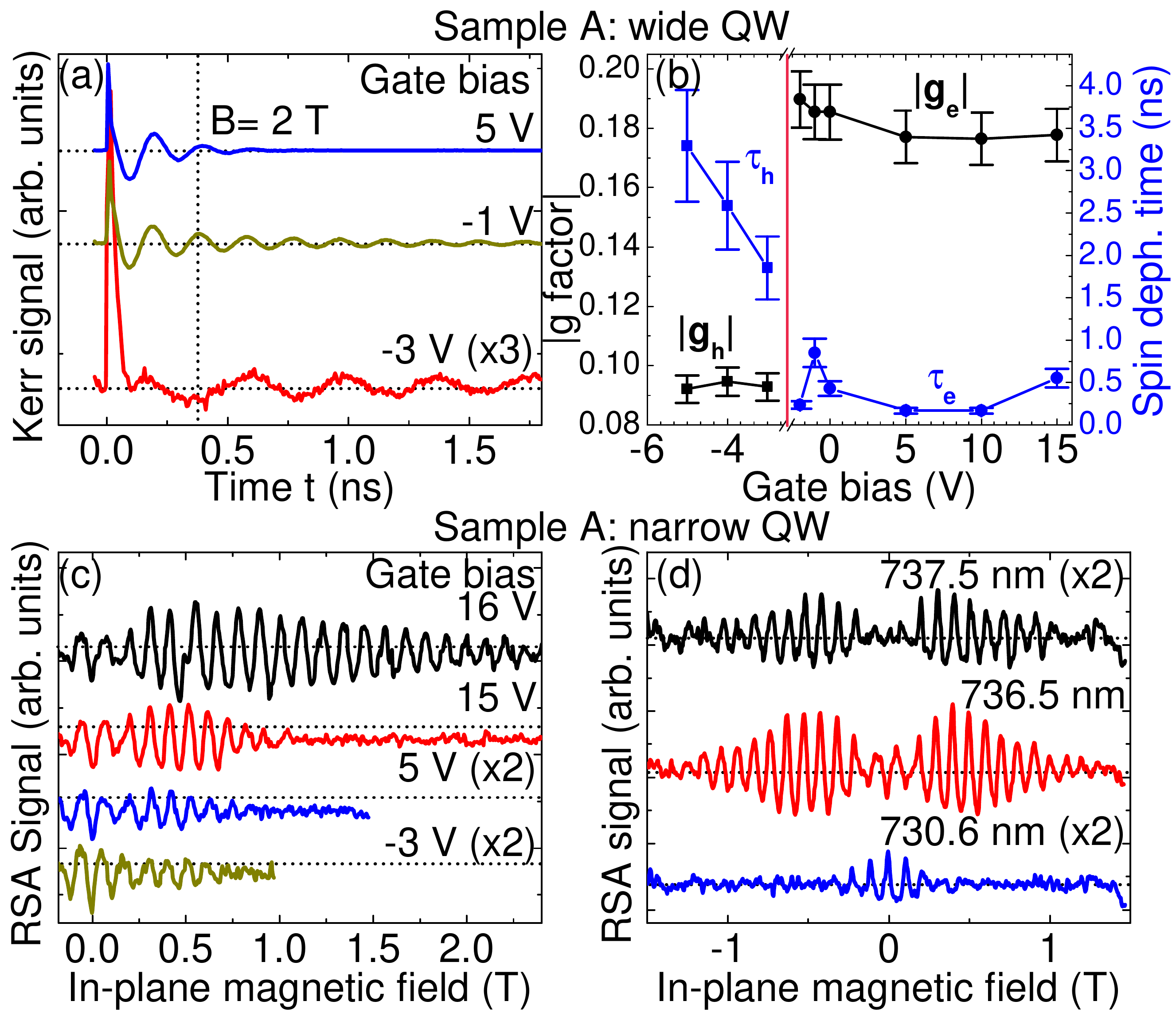}
\caption{(a) TRKR traces measured on the wide QW in sample A at a fixed  in-plane magnetic field of 2~T for different gate bias. (b) Electron/hole $g$ factors and spin dephasing times as a function of gate bias. The vertical line indicates the crossover from a hole- to an electron-dominated regime. (c) and (d) RSA measurements on the narrow QW in sample A as a function of (c) gate bias and (d) excitation wavelength.}
\label{Fig2}
\end{figure}
First,  we validate our sample design by performing time-resolved Kerr rotation (TRKR) measurements on the wide QW of the (001)-grown sample A. Figure~\ref{Fig2}(a) exemplarily shows three TRKR traces measured for different values of the gate bias with a fixed in-plane magnetic field of 2~T. All traces show a pronounced peak of the Kerr signal around zero time delay, indicating optical orientation of electron-hole pairs and partial photocarrier recombination within the QW. This is followed by an exponentially damped oscillation, which we identify as Larmor precession of the optically oriented carriers remaining in the QW. When we compare the two top traces to the bottom trace, we clearly see a large difference in the precession frequencies. We extract the spin precession frequencies and spin dephasing times (SDTs) by fitting an exponentially damped cosine function to  TRKR traces measured for a wide range of bias voltages. Whether the observed spin precession for a certain bias is caused by electrons or holes can be determined via the magnitude of the effective $g$ factor. We note that in our TRKR measurements, we cannot determine the sign of $g$.  The results are summarized in Fig.~\ref{Fig2}(b).
For bias voltages above $-3$~V, we find a nearly-constant $g$ factor value of 0.18, which we assign to electron spin precession. The  AlAs barriers in our structure lead to a stronger confinement than the more common Al$_{0.3}$Ga$_{0.7}$As barrier material, so that the e1-hh1 transition energy in our QW, measured in photoluminescence (not shown) is about 1555~meV. In a study of GaAs-Al$_{x}$Ga$_{1-x}$As QWs with different Al concentration, Yugova et al. found an effective electron $g$ factor of 0.20 for similar transition energies~\cite{yugova:245302}, in reasonable agreement with our observations.

As the bias voltage is lowered below $-3$~V, there is a sharp transition to a smaller $g$ factor of 0.09, which we  assign to hole spin precession. In contrast to previous studies on gated p-doped GaAs-AlGaAs QWs~\cite{kugler:035325}, we do not observe a pronounced dependence of the hole $g$ factor on the gate bias. This indicates that the hole wave function does not significantly penetrate into the AlAs barriers due to the large barrier height. Therefore, there is no bias-dependent admixture of the respective hole $g$ factors in the QW and barrier materials.  We note that the ensemble SDT of the holes is significantly larger than that of the electrons (see Fig.~\ref{Fig2}(b)) and increases with decreasing gate bias. Most likely, the dependence of the hole SDT on the gate bias is caused by a combination of two effects: firstly,  the tunneling rates of the holes may change with the gate bias, limiting the effective lifetime in the QW. Secondly, the position of the maximum of the hole wave function within the QW shifts with gate bias, so that the $g$ factor inhomogeneity may change and influence the ensemble SDT. This effect will be discussed below in more detail. By contrast, the electron SDT does not show a strong dependence on the bias, indicating that the  effective lifetime of electrons in the QW is limited by tunneling processes involving the X valley states in the barriers.

To explore the limits of hole spin dephasing in our sample, we perform resonant spin amplification (RSA) measurements on the narrow QW, as previous studies have shown that the large HH-LH splitting in narrow QWs leads to increased hole SDTs~\cite{Korn10}.  Figure~\ref{Fig2}(c) shows RSA traces measured for different gate biases for a laser  excitation energy slightly above the resonance of the narrow QW. In a large gate voltage range, optically oriented electrons can rapidly tunnel out of the narrow QW, as indicated in Fig.~\ref{Fig1}(b), so that excess holes remain in the QW.  Due to the interplay of hole spin relaxation,  spin-selective photocarrier recombination~\cite{Kugler_nonresonant,Salis_Hole_NonRes} and electron tunneling, we observe RSA maxima also at zero magnetic field, and a sign reversal of the RSA peaks occurs at a finite magnetic field of about 0.2~T. As the magnetic field is increased further, the amplitude of the RSA peaks decays due to hole spin ensemble dephasing caused by the hole $g$ factor inhomogeneity, $\Delta_g$~\cite{Bayer_inbook}. This leads to a characteristic magnetic-field dependence of the ensemble hole SDT, $\tau_h$:
\begin{equation}
\tau_h=\left (\frac{1}{T_2}+\frac{\Delta g_h \mu_B
B}{\hbar}\right )^{-1}.
\label{T2}
\end{equation}
Here, $T_2$ is the hole SDT in the absence of magnetic-field-induced dephasing.
Remarkably, the decay of the RSA peaks shows a strong dependence on the gate bias: for large positive bias, RSA signals can be observed up to magnetic fields of more than 2~T. To extract $T_2$ and $\Delta g_h$ from the data, we compare the RSA traces to a rate equation model~\cite{Korn10} (see appendix). We find a $T_2$ of 10~ns, and a minimum $\Delta g_h$ of 0.002. The minimum $\Delta g_h$ for our double QW structure is  below the value previously observed in p-doped QWs~\cite{Korn10,KornReview}, indicating that the absence of (modulation) doping leads to a significant reduction of local potential fluctuations. A large positive gate bias centers the hole wave function within the narrow QW, so that the effect of $g$ factor fluctuations due to interface roughness is minimized as well.

We study the initialization of the hole spin polarization in more detail by varying the laser excitation energy in the RSA measurements for a fixed gate bias. For near-resonant excitation (upper trace in Fig.~\ref{Fig2}(d)), there is only a small RSA peak observable at zero magnetic field, and the RSA amplitude builds with increasing field due to electron-precession-induced initialization of a hole spin polarization. As the excitation energy is increased (middle trace), a pronounced zero-field RSA peak is observed, which stems from an indirect initialization of the hole spin polarization. During energy relaxation, most of the optically oriented hole spins relax, while the electron spin polarization is conserved. In subsequent photocarrier recombination, spin-polarized electrons predominantly \emph{remove} holes with matching spin orientation, initializing a hole spin polarization which is oriented in the opposite direction of the optically oriented holes. As the magnetic field is increased, the precession-induced initialization becomes dominant, leading to a sign reversal of the RSA peaks. For even larger excitation energy, we observe a pronounced RSA peak at zero field, and finite-field RSA peaks with the same orientation. This shape of the RSA trace can be explained by rapid tunneling of spin-polarized electrons out of the QW, which is facilitated by the excess  energy due to nonresonant excitation. Therefore, the optically oriented hole spin polarization is not  depleted by direct photocarrier recombination at zero magnetic field. The RSA signal for these conditions can also be simulated precisely using our rate equation model (see Fig.~\ref{FigS1}(b)).
To summarize, the hole spin dynamics we  observe in our (001)-grown undoped double QW structure are in good agreement with previous results obtained on \emph{p-modulation-doped}, (001)-grown samples, demonstrating the validity of our sample design. The double QW structure  shows a smaller hole g factor inhomogeneity than modulation-doped samples due to reduced local potential fluctuations.

\subsection{In-plane hole $g$ factor anisotropy in (110)- and (113a)-grown coupled QW systems}
\begin{figure}[h]%
\centering
\includegraphics[width= \linewidth]{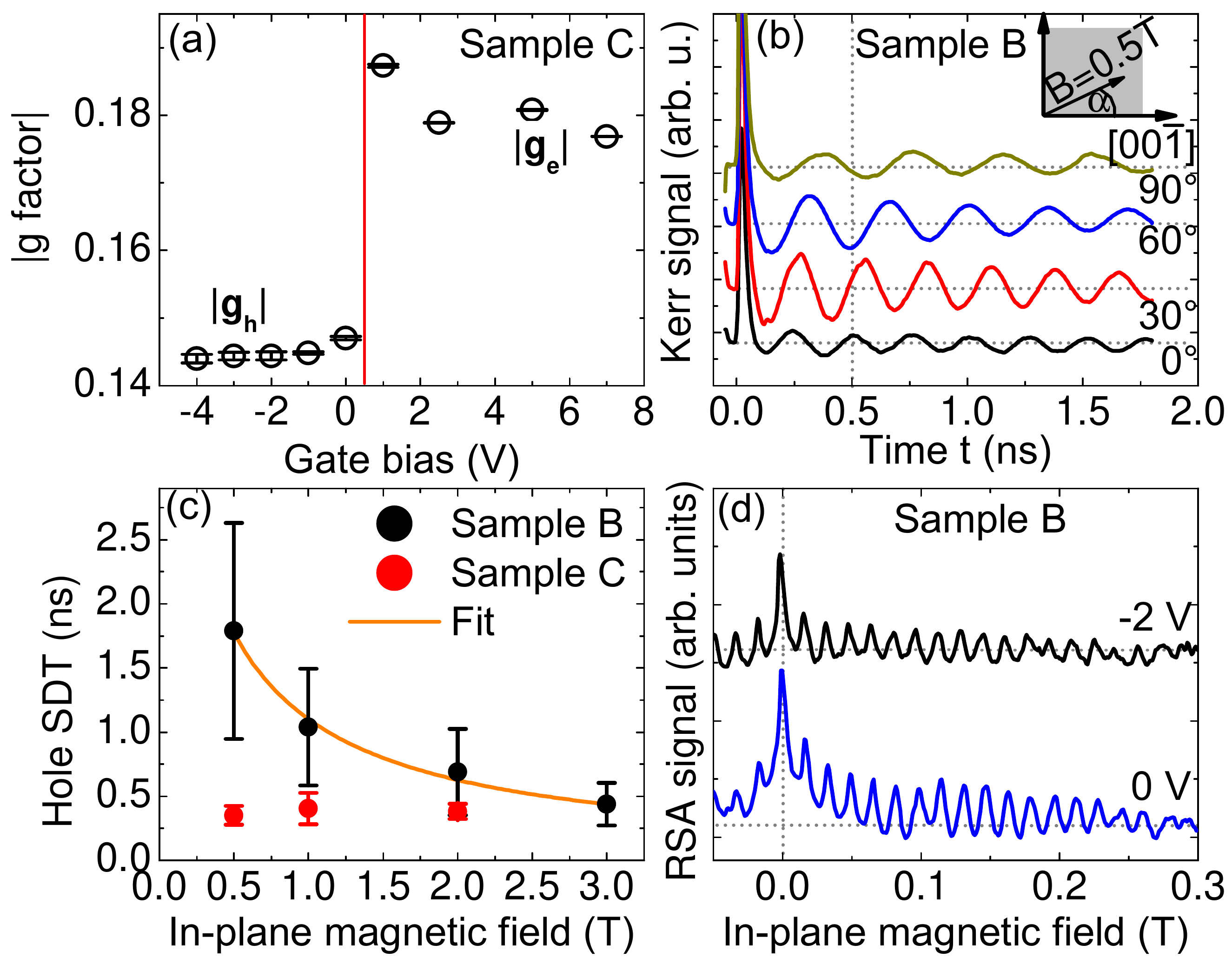}
\caption{(a) Electron/hole $g$ factors as a function of gate bias in sample C. The vertical line indicates the crossover from a hole- to an electron-dominated regime. (b) TRKR traces measured on sample B for a fixed magnitude of the in-plane magnetic field applied at various angles $\alpha$ relative to the $[00\bar{1}]$ axis. (c) Hole SDTs as a function of in-plane magnetic field for samples B and C. The data is averaged over all in-plane orientations of the samples. The solid line indicates a fit to the data using eq.~\ref{T2}.(d) RSA measurements on sample B using different gate voltages. The field is applied parallel to the $[\bar{1}10]$ axis.}
\label{Fig3}
\end{figure}

\begin{figure}[h]%
\centering
\includegraphics[width= \linewidth]{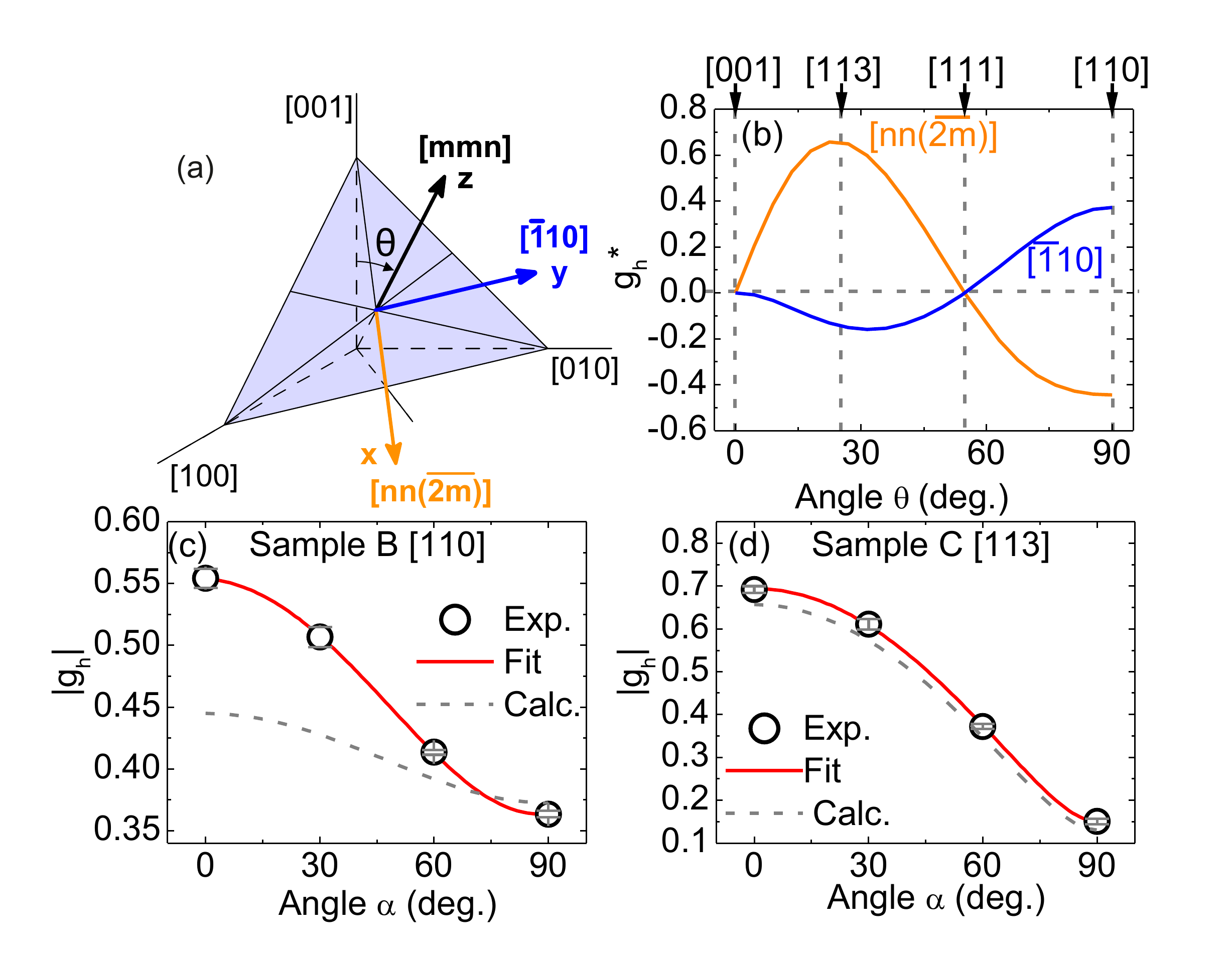}
\caption{(a)Schematic of the coordinate system used in the manuscript. (b) Numerical calculation of the hole $g$ factor for two different in-plane magnetic field directions as a function the angle between the growth axis and [001] assuming a QW width of 12~nm.   (c) and (d) Hole $g$ factor of sample B (c) and C(d) as a function of in-plane magnetic field angle $\alpha$ relative to the $[nn\overline{(2m)}]$ ($x$) axis. Solid lines in (c) and (d) indicate fits to the data using eq.~\ref{gEff}, dashed lines indicate values calculated with eq.~\ref{gEff} using the data depicted in (b).}
\label{Fig4}
\end{figure}

In our study of samples B and C, we mostly focus on TRKR measurements to determine the large hole $g$ factor values and the in-plane anisotropy predicted for (110)- and (113a)-grown QWs~\cite{winkler00}.
For both samples, we first determine the gate bias range in which a transition from electron- to hole-dominated long-lived spin precession occurs by determining the effective g factor. This is shown exemplarily for sample C in Fig.~\ref{Fig3}(a). Here, the effective g factor shows a sharp transition from a value of about 0.18 for positive gate bias, which we  assign to electron spin precession, to a value of about 0.15 as the bias is reduced to 0~V and below, corresponding to hole spin precession. Then, we perform a series of TRKR measurements with different magnetic fields for a fixed gate bias in the hole-dominated range, with the magnetic field oriented in the QW plane at an angle $\alpha$ relative to the in-plane $x$ axis.  The effective hole $g$ factor is extracted from a linear fit of the magnetic-field-dependent precession frequency. Each sample is manually mounted in four different orientations $\alpha$ relative to the magnetic field in subsequent cooling cycles. We clearly see that the precession frequency for a fixed magnetic field drastically changes with the orientation of the field relative to the crystallographic axis, as Fig.~\ref{Fig3}(b) exemplarily shows for sample B. This variation is due to the in-plane anisotropy of the hole $g$ factor. In the following we denote the in-plane crystallographic direction $[00\bar{1}]$ ($[33\bar{2}]$) of sample B (C) as $x$ axis and the direction $[\bar{1}10]$ as $y$ axis (see Fig.~\ref{Fig4}(a)). For an arbitrary angle $\alpha$ of the in-plane magnetic field relative to  the $x$ axis the effective hole $g$ factor ${g}_{h}^{*} (\alpha)$ can be calculated using the $g$ factors $g_x$ and $g_y$ in $x$ and $y$ direction:
\begin{equation}
{g}_{h}^{*} (\alpha) =
\sqrt{{g}_x^{2}cos^{2}(\alpha)
    + {g}_y^{2}sin^{2}(\alpha)}.
\label{gEff}
\end{equation}
Thus, we can extract $g_x$ and $g_y$ by fitting equation~(\ref{gEff}) to ${g}_{h}^{*} (\alpha)$ measured for each sample. The results are summarized in table~\ref{gRes}.
\begin{table}
  \centering
  \begin{tabular} {|c|c|c|}
  \hline
 Sample & axis & $\lvert g_h \rvert$ \\
  \hline
 B & $[\bar{1}10]$ & 0.364 $\pm$ 0.003 \\
  & $[00\bar{1}]$ & 0.554 $\pm$ 0.008 \\
  \hline
C &$[\bar{1}10]$ & 0.151 $\pm$ 0.007 \\
 & $[33\bar{2}]$ & 0.692 $\pm$ 0.008 \\
 \hline
\end{tabular}
  \caption{Hole $g$ factors for two in-plane crystallographic directions in samples B and C.}\label{gRes}
\end{table}

To obtain an accurate basis for comparison of our experimental results, we perform calculations of the HH Zeeman spin splitting as a function of the QW growth axis and the in-plane magnetic field orientation for a QW width of 12~nm. The coordinate system is sketched in Fig.~\ref{Fig4}(a). The growth axis in $[mmn]$ direction is rotated with respect to the [001] crystallographic axis by an angle $\theta$, such that the in-plane major directions are $[nn\overline{(2m)}]$ ($x$ axis) and $[\bar{1}10]$ ($y$ axis). Figure~\ref{Fig4}(b) shows the hole $g$ factor obtained in these calculations. While $g$ is close to zero for (001)- and (111)-grown QWs, large values (in magnitude) are obtained for (113) and (110) growth, and a pronounced anisotropy for different in-plane magnetic field orientations is clearly visible. Remarkably, a significant in-plane anisotropy is obtained here for the narrow 12~nm wide (110)-grown QW, whereas such an anisotropy is absent in previous calculations for wider (110)-grown QWs~\cite{winkler00}. By contrast, the $g$ factor for (113)-grown QWs does not vary strongly with QW width.
For sample C, the calculated values for $\lvert g_h \rvert$ as a function of the angle $\alpha$ are in very good agreement with the measured values (Fig.~\ref{Fig4}(d)). In this sample, the hole $g$ factor changes by a factor of 4 as a function of the angle $\alpha$. By contrast, we observe a smaller change of $\lvert g_h \rvert$ in sample B, and for the in-plane $[00\bar{1}]$ axis, the calculated value of $\lvert g_h \rvert$ is slightly smaller than the measured value (see Fig.~\ref{Fig4}(c)). The latter may be due to our gated sample structure, which leads to a growth-axis asymmetry of the confining potential, corresponding to a further reduction of the effective QW width.
In addition to the large in-plane anisotropy of $\lvert g_h \rvert$, there is also a weaker in-plane anisotropy of the electron $g$ factor in both samples (see Fig.~\ref{FigS3}). This anisotropy was previously observed in narrow (110)-grown QWs~\cite{Oestreich11}.
To summarize, our sample design allows us to directly observe a large in-plane anisotropy of the hole g factor for QWs grown on low-symmetry surfaces. The experimental results are in good agreement with realistic calculations of the HH Zeeman splitting that take the width of our QW structures into account.

\subsection{Hole spin dephasing anisotropy in (110)-grown coupled QW system}
We investigate the magnetic-field dependence of the hole SDT in samples B and C to determine the mechanism limiting hole spin coherence. For this, we average the hole SDT values measured by TRKR for a certain magnetic field over all magnetic field orientations. The results are depicted in Fig.~\ref{Fig3}(c), demonstrating that the two samples show rather different behavior: in sample B, the hole SDT shows a characteristic $B^{-1}$ dependence as described by eq.~\ref{T2}, indicating that $g$ factor inhomogeneity limits the hole SDT in high magnetic fields.  By contrast, the hole SDT in sample C is significantly shorter (below 400~ps), and independent of the magnetic field. In this sample, carrier tunneling and recombination may limit the effective hole SDT.
To study the hole SDT in low magnetic fields, we also perform RSA measurements on sample B. Figure~\ref{Fig3}(d) shows RSA traces, measured with the magnetic field applied parallel to the $[\bar{1}10]$ axis, for different applied gate voltages in the range where we observe hole spin dynamics in TRKR. For the gate voltages investigated, we find a pronounced RSA maximum for $B=0$~T, with all finite-field RSA maxima having the same orientation. This shows that the long-lived hole spin polarization is initialized directly by the optically oriented holes, while spin-polarized electrons rapidly tunnel out of the QW, so that there is no complex interplay between electron and hole spin dynamics. Remarkably, we find that the RSA maximum amplitude for zero magnetic field is about two times larger than that of the first finite-field maximum, while the amplitudes of subsequent finite-field maxima decrease slowly, indicating weak ensemble dephasing. The shape of this RSA trace cannot be described within our rate equation model (see Fig.~\ref{FigS1}(b)). It may be explained by an orientational anisotropy of the hole spin dephasing, with a significantly larger hole SDT for spins oriented along the QW normal. Such an anisotropy was previously observed for \emph{electrons} in (110)-grown QWs~\cite{Griesbeck12}. To our knowledge, the influence of the QW symmetry on \emph{hole} spin dephasing has only been considered for (111)-grown QWs, so far~\cite{Wu_Hole_12}, and merits further investigation.
\section*{Conclusion}
In summary, we have studied spin dynamics in undoped double QW systems. We find that  long-lived spin dynamics  can be observed in these structures due to a spatial separation of electrons and holes. Gate-dependent measurements demonstrate that either long-lived electron- or hole-spin polarization can be generated in one of the QWs. In resonant spin amplification measurements, we demonstrate hole spin dephasing times of up to 10~ns and a very low inhomogeneity of the hole $g$ factor. The fact that our  structures are undoped allows us to prepare samples grown along different crystallographic orientations. In (110)- and (113a)-grown structures, we find a large in-plane anisotropy of the hole $g$ factor, in agreement with our numerical calculations, and a weaker in-plane anisotropy of the electron $g$ factor. In the (110)-grown sample, we find indications of anisotropic hole spin dephasing.  Our results may pave the way for future studies on the control of hole spin dynamics using crystal symmetry engineering.

\section*{Acknowledgements}
Financial support by the DFG via SPP 1285 and SFB 689 is gratefully acknowledged.
\section*{Appendix}
\subsection*{Extraction of spin dynamics parameters from RSA traces}
\begin{figure}
\centering
\includegraphics[width= 0.8 \linewidth]{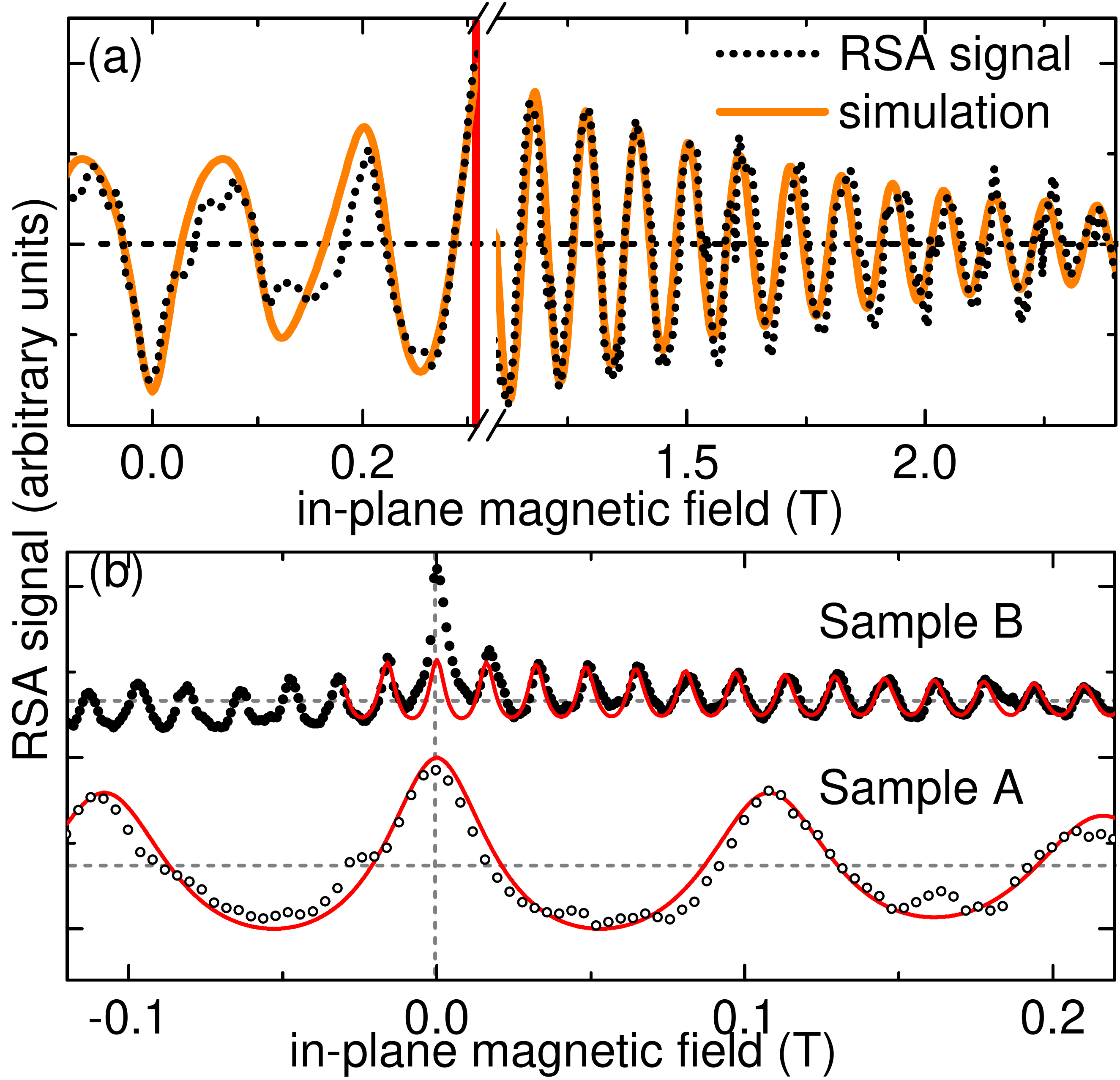}
\caption{(a) Low- and high-field regions of RSA trace measured on the narrow QW in sample A for 16~V gate bias compared to simulated RSA trace. (b) RSA traces measured on  sample B (top trace, dots) and sample A (bottom trace, open circles)  compared to simulated RSA traces (solid lines).}
\label{FigS1}
\end{figure}
The characteristic shape of the RSA traces due to hole spin dynamics has been previously studied in p-doped QWs~\cite{Korn10,KornReview}. It arises from the interplay of spin and photocarrier dynamics. Under resonant optical excitation of a QW, oppositely equal electron and hole spin polarizations are created, and in the absence of a magnetic field or other processes which would create an imbalance between electron and hole spin polarizations, spin-polarized electrons recombine with holes that match their spin orientation, so that photocarrier recombination removes the spin polarizations, and no transfer of hole spin polarization to resident carriers occurs. Therefore, in measurements under these conditions, no RSA peak at zero magnetic field is observed. For finite magnetic fields, which lead to different spin precession frequencies for electrons and holes, some transfer of spin polarization to resident holes occurs, and RSA peaks are observed.  Nonresonant excitation leads to a partial dephasing of the optically oriented hole spin polarization during energy relaxation. During photocarrier recombination, optically oriented electrons preferably remove hole spins with matching orientation, so that an excess of hole spins with opposite spin orientation remains in the sample. This is visible as a zero-field RSA peak with orientation opposite to finite-field RSA peaks.
The combined dynamics of the electron and hole spin polarizations can be described via coupled equations of motion for the electron (\textbf{e}) and hole(\textbf{h}) spin polarization vectors:
\begin{eqnarray}
\nonumber  \frac{d \textbf{e}}{dt} &=& -\frac{\textbf{e}}{\tau_R} + \frac{g_e \mu_B}{\hbar} (\textbf{B}\times\textbf{e})\\
\nonumber  \frac{d \textbf{h}}{dt} &=& -\frac{\textbf{h}}{\tau_h}  + \frac{g_h \mu_B}{\hbar} (\textbf{B}\times\textbf{h}) + \frac{e_z \textbf{z}}{\tau_R},
\end{eqnarray}
with $\tau_R$ being the photocarrier recombination time.
Here, the effect of a fast partial hole spin dephasing during energy relaxation can be modeled by using different initial values for electron and hole spin polarizations.  To include the effects of ensemble dephasing due to $g$ factor inhomogeneity~\cite{Bayer_inbook}, we utilize the magnetic-field dependence of $\tau_h$ described by eq.~\ref{T2}.
We apply this model to extract the hole spin dynamics parameters  for the RSA measurements performed on sample A. Fig.~\ref{FigS1}(a) shows the best fit to the experimental data for near-resonant excitation and a gate voltage of 16~V (top trace in Fig.~\ref{Fig2}(c)) in the low and high magnetic field ranges. We find $g_h = 0.051$, $\Delta g_h = 0.002$, and $T_2 = 8$~ns. The same model may also be used to simulate RSA traces measured on sample A under highly nonresonant excitation conditions, as  Fig.~\ref{FigS1}(b) demonstrates. Under these conditions, spin-polarized electrons rapidly tunnel out of the QW, so that we only need to consider hole spin precession and dephasing. We find $g_h = 0.0527$, $\Delta g_h = 0.0022$, and $T_2 = 10$~ns. With the same approach, we are able to simulate the RSA traces measured on sample B, with the exception of the zero-field RSA maximum.
\subsection*{Electron $g$ factor anisotropy}
\begin{figure}
\centering
\includegraphics[width= 0.8 \linewidth]{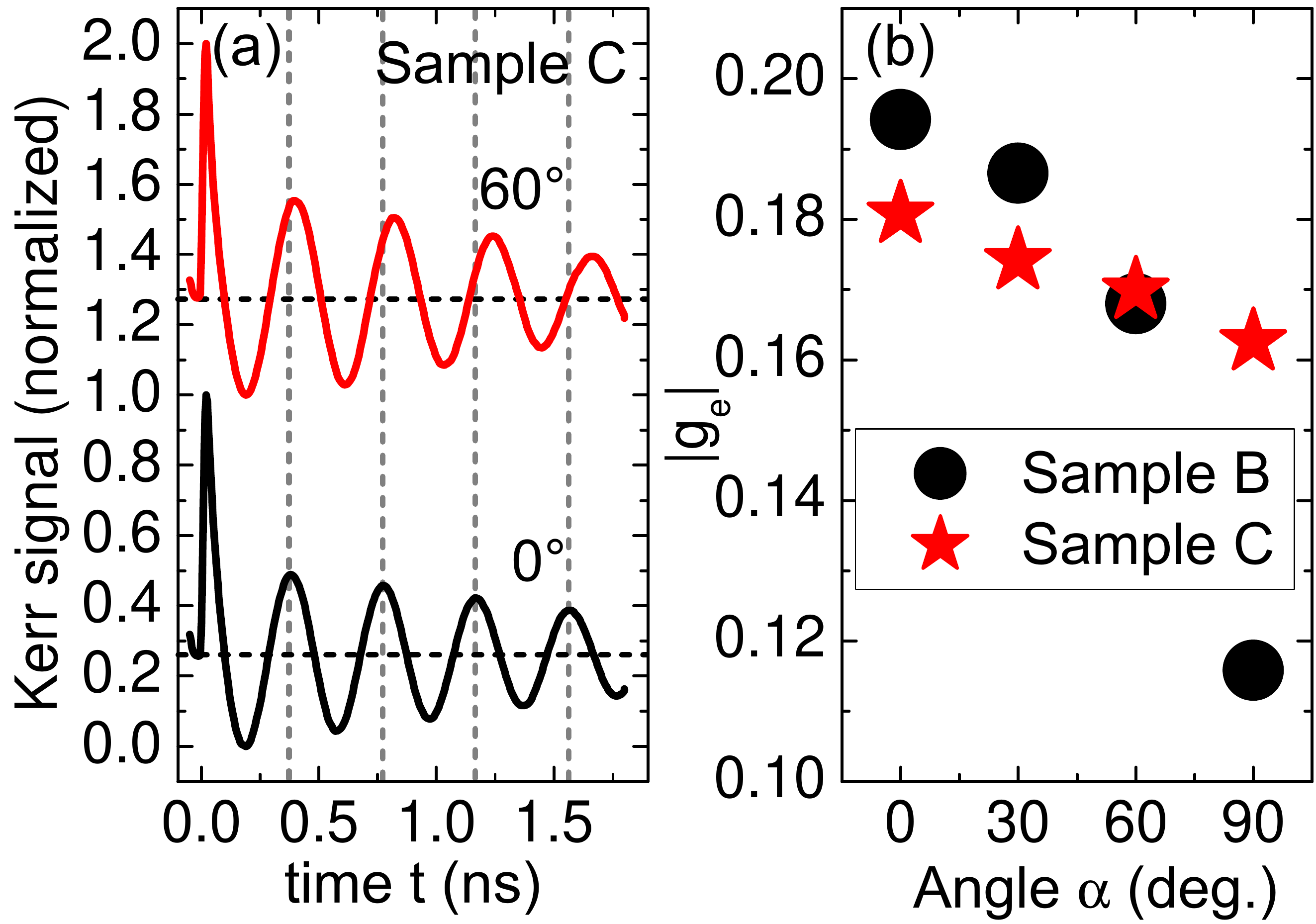}
\caption{(a) TRKR traces measured on the wide QW in sample C at a fixed magnetic field B = 1~T for two different in-plane orientations of the field. The vertical lines serve as guide to the eye. (b) Electron $g$ factors as a function of in-plane magnetic field angle for samples B and C. }
\label{FigS3}
\end{figure}
We investigate the in-plane anisotropy of the electron $g$ factor in samples B and C by studying spin precession in the wide QW for fixed gate voltages in the range where long-lived electron spin precession is observable. In both samples, we find a systematic variation of the electron $g$ factor with the in-plane magnetic field orientation. As Fig.~\ref{FigS3} shows, this anisotropy is significantly more pronounced in sample B. Recently, in a study of undoped, (110)-grown QWs with AlGaAs barriers, a similar in-plane anisotropy of the electron $g$ factor was observed~\cite{Oestreich11} and shown to originate from the low-symmetry growth orientation (110).

\end{document}